\documentclass[10pt,letterpaper]{article}
\usepackage{opex3}
\usepackage{cite}

\begin{document}




\title{Electronic control of extraordinary terahertz transmission through subwavelength metal hole arrays}

\author{Hou-Tong Chen$^{1}$, Hong Lu$^2$, Abul K. Azad$^1$, Richard D. Averitt$^3$, Arthur C. Gossard$^2$, Stuart A. Trugman$^1$, John F. O'Hara$^1$, and Antoinette J. Taylor$^1$}

\address{$^1$Los Alamos National Laboratory, MPA-CINT, MS K771, Los Alamos, New Mexico 87545, USA}
\address{$^2$Materials Department, University of California, Santa Barbara, California 93106, USA}
\address{$^3$Department of Physics, Boston University, 590 Commonwealth Avenue, Boston, Massachusetts 02215, USA}
\email{chenht@lanl.gov} 



\begin{abstract}
We describe the electronic control of extraordinary terahertz transmission through subwavelength metal hole arrays fabricated on doped semiconductor substrates. The hybrid metal-semiconductor forms a Schottky diode structure, where the active depletion region modifies the substrate conductivity in real-time by applying an external voltage bias. This enables effective control of the resonance enhanced terahertz transmission. Our proof of principle device achieves an intensity modulation depth of 52\% by changing the voltage bias between 0 and 16 volts. Further optimization may result in improvement of device performance and practical applications. This approach can be also translated to the other optical frequency ranges.
\end{abstract}

~



\section{Introduction}
Artificially structured composite materials are playing an increasingly important role in overcoming the deficiency of natural materials to obtain a functional response in the terahertz (THz) frequency range. One excellent example is the successful demonstration of THz quantum cascade lasers (QCLs) using composite semiconductor heterostructures~\cite{Kohler2002,Rochat2002,Williams2003}. The so-called ``THz gap'' has resulted in the general failure to translate technologies at microwave and optical frequencies to the THz frequency range~\cite{Ferguson2002}. The recent progress in THz metamaterials~\cite{Yen2004,Padilla2006,Chen2006,Chen2007B,Chen2007}, photonic crystals~\cite{Nemec2004,Nemec2005,Fekete2007,Savelev2005,Savelev2006}, and subwavelength metallic hole arrays~\cite{Rivas2003,Qu2004,Cao2004,Tanaka2005,Matsui2007,Zhang2007}, also proposed or demonstrated high performance functional THz devices, which may ultimately result in a complete manipulation of THz waves.

Since their experimental demonstration in the optical frequency range~\cite{Ebbesen1998}, subwavelength metal hole arrays have attracted considerable attention regarding extraordinary optical transmission at THz frequencies~\cite{Qu2004,Cao2004,Tanaka2005,Matsui2007}. Such phenomena are generally attributed to the resonant excitation of surface plasmon (SP) modes at a metal-dielectric interface~\cite{Ebbesen1998,Ghaemi1998}. In the THz frequency range, the dielectric constant of metals exhibits very large values, thus the free-space wavelength of the fundamental SP mode excited by the normally incident electromagnetic waves in a square array of metal holes can be approximated as~\cite{Qu2004,Ghaemi1998}, 
\begin{equation}
\lambda_{res}=L \sqrt{\epsilon_1},
\end{equation}
where $L$ is the lattice parameter of the metal hole array and $\epsilon_1$ is the dielectric constant of the interface medium. The enhanced transmission exceeds the geometrical transmission~\cite{Ebbesen1998}, and is orders of magnitude higher than that expected for non-resonant transmission through sub-wavelength metal holes~\cite{Bethe1944}.

The majority of research has focused on the passive properties of extraordinary THz transmission, e.g., the effects of metal film thickness~\cite{Azad2005}, hole geometry~\cite{Lee2006}, periodicity~\cite{Matsui2007}, etc. It has been shown that the resonance can also be affected by depositing a dielectric layer on the metal hole arrays~\cite{Tanaka2005} and by doping a semiconductor substrate~\cite{Wasserman2007}, both of which result in significant shifting of the resonance frequency. However, little work has focused on the active manipulation of the extraordinary optical transmission though it is essential to realize many applications. All-optical switching and modulation were demonstrated~\cite{Krasavin2004,Janke2005,Dintinger2006,Zhang2007} by stimulated optical modification of either the metallic surface or the interface dielectric medium. Liquid crystal~\cite{Kim1999} and thermal-optical~\cite{Nikolajsen2004,Park2006} approaches were also used to switch and modulate optical signals via the application of electric field or electrical heating; however, these inherently slow operations severely restrict potential applications. Additionally most of these demonstrations are in the visible. In this paper, we demonstrate electronic switching of the extraordinary THz transmission through subwavelength metal hole arrays fabricated on doped semiconductor substrates. The passive resonance properties are mainly determined by the geometry and dimensions of the metal holes as well as the array periodicity. By electronically altering the substrate conductivity via an external voltage bias, we accomplish switching of the extraordinary THz transmission in real time with an intensity modulation depth as high as 52\%, which is comparable with the electronic THz modulation using metamaterial devices~\cite{Chen2006}.

\section{Sample design and fabrication}
\begin{figure}[!t]
\begin{center}
\begin{tabular}{c}
\centerline{\includegraphics[width=4in]{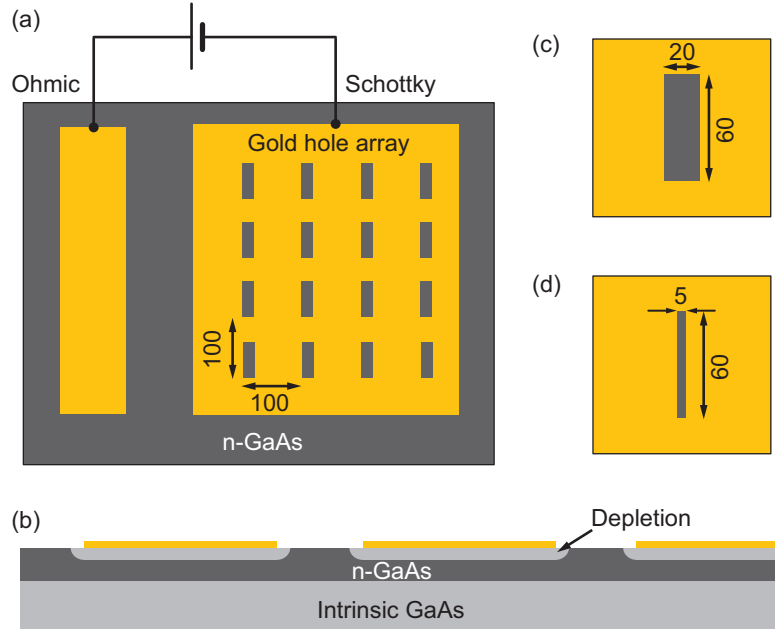}}
\end{tabular}
\end{center}
\caption
{(a) Schematic design of the metal hole arrays exhibiting an electronically switchable extraordinary THz transmission. (b) Cross-sectional view of structures showing the depletion regions under reverse voltage bias. Two unit cells are shown with (c) wide and (d) narrow holes. All dimensions are in microns.}
\label{Fig1}
\end{figure}

In Fig.~\ref{Fig1}(a) we show the conceptual design of subwavelength metal hole arrays for the electronically switchable resonance enhanced extraordinary THz transmission. The \textit{n}-doped semiconductor and metal film form a Schottky diode structure, where the depletion region can be actively controlled with an external voltage bias, thereby switching the damping of the resonance and the extraordinary THz transmission. The samples characterized here were fabricated on gallium arsenide (GaAs) substrates. The substrates have a 2~$\mu$m thick \textit{n}-doped GaAs layer with a free carrier density of $3.2 \times 10^{16}$~cm$^{-3}$ on an intrinsic GaAs wafer grown by molecular beam epitaxy. An ohmic contact surrounding but separated from the hole array of approximately 1~mm was fabricated by electron beam deposition of 20~nm of germanium, 20~nm of gold, 20~nm of nickel, and 200~nm of gold in sequence, followed by rapid thermal annealing at 400$^\circ$C for 1 minute. The metal hole arrays were fabricated using standard photolithographic methods and electron beam deposition of 10~nm of titanium and 200~nm of gold, followed by a lift-off process. Two samples with hole geometries shown in Figs.~\ref{Fig1}(c) and (d) were made. The hole widths are 20~$\mu$m and 5~$\mu$m, respectively. The same metal hole arrays were also fabricated on intrinsic GaAs substrates to enable a comparison of the resonance damping caused by the lossy \textit{n}-doped GaAs substrates.

\section{Experiments and results}
\begin{figure}[!t]
\begin{center}
\begin{tabular}{c}
\centerline{\includegraphics[width=3in]{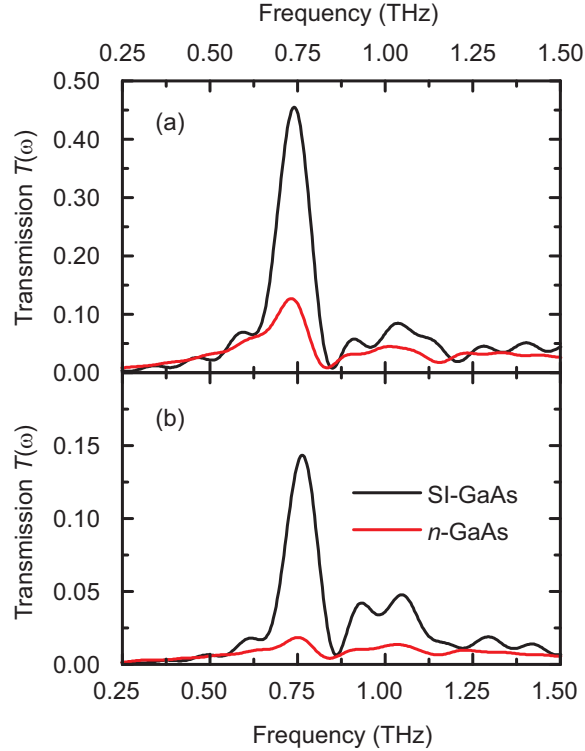}}
\end{tabular}
\end{center}
\caption
{THz intensity transmission spectra for samples fabricated on an intrinsic GaAs substrate (black curves) and on an \textit{n}-GaAs substrate (red curves). (a) Hole width 20~$\mu$m and (b) hole width 5~$\mu$m.}
\label{Fig2}
\end{figure}

We measured the frequency dependent THz transmission using THz time-domain spectroscopy (THz-TDS)~\cite{Grischkowsky1990}. The electric field of the THz pulses was coherently measured in the time-domain after propagating through the metal hole array samples and a bare substrate serving as the reference. Fourier transformation of the time-domain data then yielded the frequency dependent THz electric field amplitude and phase. Dividing the sample's complex spectrum by the reference we obtained the normalized transmission amplitude $t(\omega)$ and phase $\phi(\omega)$ through the metal hole arrays. The intensity (or power) transmission is given by $T(\omega)=t^2(\omega)$. The THz time-domain data was temporally windowed to eliminate effects of multiple reflections within the GaAs substrate. This, however, truncates the THz oscillation in the time-domain data yielding ringing structures in the transmission spectra, particularly for the metal hole arrays fabricated on intrinsic GaAs substrates where the damping is small, as shown by the black curves in Fig.~\ref{Fig2}.

Without any applied voltage bias, the measured THz transmission is shown in Fig.~\ref{Fig2} for the metal hole arrays fabricated on \textit{n}-doped GaAs and intrinsic GaAs substrates, respectively. The intrinsic GaAs substrate can be considered as a lossless dielectric material at THz frequencies. We observed transmission peaks near 0.75~THz and minima (Wood's anomalies) at 0.84~THz for the lowest modes as shown by the black curves in Fig.~\ref{Fig2}, which are as expected and whose frequencies are determined by the periodicity of metal hole arrays. The values of the peak intensity transmission are 4 and 5 times as large as those of the geometrical transmission (i.e. the fraction of hole area normalized to the unit cell) for sample geometries in Figs.~\ref{Fig1}(c) and (d), respectively. When \textit{n}-doped GaAs substrates are used, the extraordinary THz transmission is significantly damped as shown by the red curves in Fig.~\ref{Fig2}. The values of peak intensity transmission are only 1 and 0.6 times as large as those of the geometrical transmission for sample geometries in Figs.~\ref{Fig1}(c) and (d), respectively. In addition, we observed a small red-shift of THz transmission peaks and Wood's anomalies, which is consistent with results in the mid-infrared frequency range where metal hole arrays were fabricated on GaAs substrates with various doping concentration~\cite{Wasserman2007}. This shift is associated with the change of dielectric constant in the 2~$\mu$m \textit{n}-doped GaAs layer, and is confirmed by finite-element numerical simulations.

\begin{figure}[!t]
\begin{center}
\begin{tabular}{c}
\centerline{\includegraphics[width=3in]{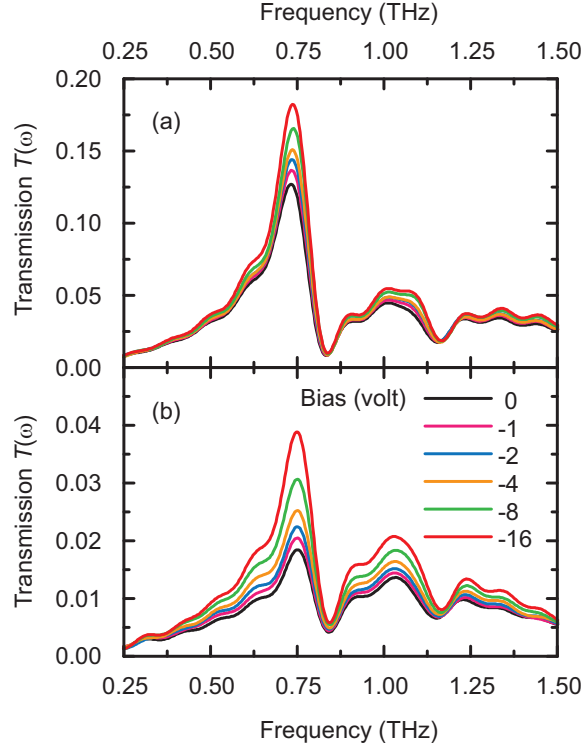}}
\end{tabular}
\end{center}
\caption
{The THz intensity transmission spectra as a function of the applied reverse voltage bias for samples fabricated on \textit{n}-GaAs substrates. (a) Hole width 20~$\mu$m and (b) hole width 5~$\mu$m.}
\label{Fig3}
\end{figure}

As described above, the substrate charge carrier density and conductivity of the 2~$\mu$m \textit{n}-doped GaAs layer can be electronically modified by applying a reverse voltage bias to the Schottky diode structure. Higher reverse voltage bias increases the depletion region thereby reducing the damping of the resonance. In Figs.~\ref{Fig3}(a) and (b) we show the experimental THz transmission spectra as a function of the reverse voltage bias for the metal hole arrays fabricated on \textit{n}-doped GaAs substrates and shown in Figs.~\ref{Fig1}(c) and (d), respectively. In both samples we observed increasing values of THz transmission peaks as the applied reverse voltage bias increased, while the transmission dips are much less affected. Furthermore, the results reveal that the modulation depth of THz transmission is also dependent on the dimensions of the rectangular holes. Under reverse voltage biases of 0 and 16 volts, the intensity modulation depth of the transmitted THz radiation is defined as $h = (T_{16V} - T_{0V})/T_{16V}$. For the sample with the geometry in Fig.~\ref{Fig1}(c) the modulation depth is $h_c = 30\%$, while it is as high as $h_d = 52\%$ for the sample having narrower metal holes in Fig.~\ref{Fig1}(d).

\section{Discussion}
The above experimental results show that the resonance strength is very sensitive to the substrate conductivity and loss. Without the voltage bias, the conducting substrate provides a large loss, as well as a ``short-circuit'' of the metal holes, thereby significantly damping the resonance and reducing the THz transmission. With the reverse voltage bias, on the other hand, the increasing depletion reduces the substrate loss, thereby enhancing the resonance and extraordinary THz transmission. Furthermore, the depletion happens not only in the regions underneath metal, but also extends laterally to the hole regions as indicated in Fig.~\ref{Fig1}(b). Since the lateral extension of the depletion region into the holes is the same for the wider and narrower holes under the same voltage bias, this results in a larger proportion of depletion area in the narrower holes. On the other hand, the electromagnetic field is significantly enhanced in the hole regions at resonance, which has been confirmed by NSOM measurements~\cite{Ghaemi1998} and numerical simulations. This means that the loss in the hole regions plays an important role in switching the extraordinary THz transmission. In the sample with narrower holes, a larger portion of depletion has more impact on reducing the damping of the resonance yielding a higher modulation depth of the extraordinary THz transmission, as shown in Fig.~\ref{Fig3}. Further narrowing of the rectangular holes should yield even higher modulation depths at the expense of lower total transmission efficiency. The depletion dependence of the resonance and extraordinary optical transmission may be even more promising for shorter wavelengths, where the required dimensions of individual holes are smaller and the hole area could be completely depleted, producing a larger total transmission through the sample.

In semiconductor substrates, the dielectric constant varies with doping concentration. According to Eq.~(1) the resonance frequency will therefore experience a significant shift~\cite{Wasserman2007}. However, with the reverse voltage bias, which should significantly change the dielectric constant of the 2~$\mu$m thick \textit{n}-doped GaAs layer, we observed only a small frequency shift as shown in Fig.~\ref{Fig3}. The explanation is that, due to the long wavelength of THz radiation, its decay into the substrate could be as large as hundreds of micrometers. So modification of the dielectric constant in the 2~$\mu$m thick \textit{n}-doped GaAs layer by applying the reverse voltage bias only has a very small effect on the resonance frequency.  

Finally, we exclude the possibility of thermal effects from the current flow under the reverse voltage bias by performing measurements for the samples under forward voltage bias, where the change of depletion is negligible but the current flow is much larger than under reverse bias. We observed no perceptible change in the THz transmission as compared to the unbiased transmission (see Fig. 4), thus the current flow or sample heating has a negligible effect on device performance. The switching of extraordinary THz transmission is purely due to the electronic modification of the substrate conductivity and loss in the 2~$\mu$m thick \textit{n}-doped GaAs layer. 

\begin{figure}[!t]
\begin{center}
\begin{tabular}{c}
\centerline{\includegraphics[width=3in]{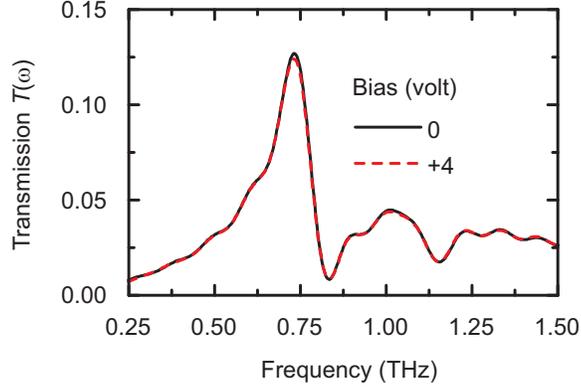}}
\end{tabular}
\end{center}
\caption
{The THz intensity transmission spectra at 0 (black solid curve) and 4 volts (red dashed curve) forward voltage bias.}
\label{Fig4}
\end{figure}

\section{Conclusion}
In conclusion, we have demonstrated the electronically switchable resonance enhanced extraordinary optical transmission at THz frequencies. The formation of a Schottky diode structure between the metal film and doped semiconductor substrate enables real-time modification of the substrate conductivity and loss, resulting in switching of the resonance and extraordinary THz transmission. The depletion extends into the hole regions, so that the metal hole geometry and dimensions play an important role, because at resonance the electromagnetic field is significantly enhanced in the metal hole regions. Further, a metal hole array with smaller hole dimensions enables a higher modulation depth. In our first generation proof-of-principle devices, we accomplished an intensity modulation depth as high as 52\%. Optimizing the hole geometry and dimensions as well as the free carrier concentration and thickness of the doped semiconductor substrate layer, the device performance may be further improved. Due to the long decay length of the THz field into the substrate, the change of dielectric constant with increasing depletion in the thin doped semiconductor layer is insufficient to significantly tune the resonance frequency. These results are scalable to higher optical frequencies, where the switching performance could be even more promising.  

\section*{Acknowledgements}
We acknowledge support from the Los Alamos National Laboratory LDRD program. This work was performed, in part, at the Center for Integrated Nanotechnologies, a U.S. Department of Energy, Office of Basic Energy Sciences Nanoscale Science Research Center operated jointly by Los Alamos and Sandia National Laboratories. Los Alamos National Laboratory, an affirmative action/equal opportunity employer, is operated by Los Alamos National Security, LLC for the National Nuclear Security Administration of US Department of Energy under contract DE-AC52-06NA25396.
\end{document}